\documentclass[]{spie}  

\usepackage{amsmath,amsfonts,amssymb}
\usepackage{graphicx}
\usepackage[colorlinks=true, allcolors=blue]{hyperref}

\title{Proof of concept of the COSMOCal project at the IRAM 30m telescope}

\author[a$^{\dagger}$]{S.~Savorgnano}
\author[b]{L.~Bizzarri}
\author[c]{A.~Ritacco}
\author[d]{J.~Aumont}
\author[e]{F.~Boulanger}
\author[c]{A.~Catalano}
\author[f]{F.~Cuttaia}
\author[e]{A.~Denis}
\author[g]{F.-X.~Désert}
\author[h]{D.~González Ovejero}
\author[i]{S.~Leclercq}
\author[c]{J.-F.~Macías-Pérez}
\author[j]{B.~Maffei}
\author[k]{M.~Migliaccio}
\author[d]{L.~Montier}
\author[e]{P.~Morfin}
\author[j]{L.~Mousset}
\author[l]{M.~Murgia}
\author[b]{F.~Nati}
\author[l]{P.~Ortu}
\author[e]{M.~Pérault}
\author[m]{G.~Pisano}
\author[l]{T.~Pisanu}
\author[g]{N.~Ponthieu}
\author[f]{L.~Terenzi}
\author[n]{J.~Treuttel}
\author[d]{L.~Vacher}
\author[b]{M.~Zannoni}

\affil[a]{Department of Physics, Boston University, Boston, MA, USA}
\affil[b]{Dipartimento di Fisica, Università di Milano-Bicocca, Milano, Italy}
\affil[c]{Université Grenoble Alpes, CNRS, LPSC-IN2P3, Grenoble, France}
\affil[d]{Institut de Recherche en Astrophysique et Planétologie (IRAP), CNRS, France}
\affil[e]{Laboratoire de Physique de l'École Normale Supérieure, ENS, Université PSL, CNRS, Sorbonne Université, Université Paris Cité, Paris, France}
\affil[f]{INAF -- Osservatorio di Astrofisica e Scienza dello Spazio di Bologna, Bologna, Italy}
\affil[g]{Univ. Grenoble Alpes, CNRS, IPAG, Grenoble, France}
\affil[h]{Institut d'Électronique et des Technologies du Numérique (IETR), France}
\affil[i]{Institut de Radioastronomie Millimétrique (IRAM), France}
\affil[j]{Institut d'Astrophysique Spatiale (IAS), CNRS, Université Paris-Saclay, France}
\affil[k]{Dipartimento di Fisica, Università di Roma Tor Vergata, Rome, Italy}
\affil[l]{INAF -- Osservatorio Astronomico di Cagliari, Cagliari, Italy}
\affil[m]{Sapienza Università di Roma, Rome, Italy}
\affil[n]{LERMA, Observatoire de Paris-PSL, CNRS, France}

\authorinfo{$^{\dagger}$Corresponding author: ssavorgn@bu.edu}

\begin{document} 
\maketitle

\begin{abstract}
This work reports on a test campaign conducted at the IRAM 30m telescope to validate the COSMOCal instrument, a novel concept for absolute calibration of polarization angle, beam properties, and instrumental efficiency. The development is motivated by the stringent calibration requirements of current and next-generation Cosmic Microwave Background (CMB) experiments, whose goal is the precise measurement of CMB polarization. Such measurements are essential to probe fundamental physics, including the possible imprint of primordial inflation. The COSMOCal\footnote{Project's website: \url{https://sites.google.com/view/cosmocal-website/home-page?authuser=0}.} concept is designed as a space-borne calibration reference intended to provide a stable and absolute polarized signal observable simultaneously by multiple large-aperture ground-based observatories. Prior to any space deployment, a terrestrial prototype was developed and characterized in the laboratory, as reported in Ritacco et al. 2024\cite{ritacco2024}. The results presented here describe the first validation of the system coupled to an antenna, carried out at the IRAM 30m telescope in September 2024. This campaign aimed at assessing the performance of the full instrument under realistic observational conditions, including its ability to reconstruct polarization observables and interface with a state-of-the-art millimeter-wave polarimeter.
\end{abstract}

\section{Scientific purpose}
Future CMB polarization experiments require an unprecedented accuracy in the calibration of the absolute polarization angle in order to achieve their scientific objectives, including the detection of primordial gravitational waves through the tensor-to-scalar ratio $r$ and the search for cosmic birefringence (Rosset et al. 2010~\cite{rosset2010}, Kaufman et al. 2014~\cite{kaufman2014}, Ritacco et al. 2022~\cite{ritacco2022}, Aumont et al. 2020~\cite{aumont2020}, Louis et al. 2025~\cite{louis2025}). Although every instrument undergoes extensive laboratory calibration before deployment, an in-flight verification of its polarization properties is essential once integrated with the telescope. Existing astronomical calibrators, such as the Crab Nebula (Aumont et al. 2010~\cite{aumont2010}, Ritacco et al. 2018~\cite{ritacco2018}), provide only a relative reference, as their measured polarization angle is more likely affected by the systematic uncertainties of the instrument used to observe them. Consequently, they cannot provide the absolute calibration accuracy required by next-generation CMB experiments.

To overcome this limitation, several artificial calibration strategies have been proposed, including ground-based or drone-mounted polarized sources (Coppi et al. 2025\cite{coppi2025}, Nati et al. 2017~\cite{nati2017}, Cornelison et al. 2022~\cite{cornelison2022}) and a dedicated CubeSat at the L2 Lagrange point for a single CMB mission (Casas et al. 2021~\cite{casas2021}). While these concepts are well suited to small-aperture telescopes or individual space missions, they cannot provide a common far-field calibrator for large-aperture ground-based observatories, whose far-field distance exceeds the Earth's atmosphere. This motivates the development of COSMOCal (Ritacco et al. 2024\cite{ritacco2024}), which will place a polarized microwave calibration source in geostationary orbit. From this location, the source will be continuously visible from both Europe and Chile, providing a stable far-field reference for a broad range of telescopes.

The payload will emit linearly polarized monochromatic signals at 90, 150 and 270~GHz, with an absolute polarization angle accuracy better than $0.1^\circ$. The 90 and 150~GHz channels directly address the primary observing bands of current and future CMB experiments, while the addition of the 270~GHz channel extends the calibration to frequencies where polarized Galactic dust emission becomes dominant. This channel is particularly important for improving our understanding of dust polarization systematics and for validating foreground-removal strategies that are essential for future CMB analyses.

The initial observatories targeted by COSMOCal are the Simons Observatory Large Aperture Telescope (LAT), the Sardinia Radio Telescope (SRT), and the IRAM 30m telescope. In particular, the inclusion of the IRAM 30m telescope enables, for the first time, an absolute polarization calibration at 270~GHz. Although the 30m telescope is not designed to directly constrain Galactic dust foregrounds over large sky areas, it will provide highly accurate observations of polarized calibration standards, such as the Crab Nebula, establishing reference measurements for cross-calibrating other instruments. At the same time, COSMOCal will broaden the scientific capabilities of the IRAM 30m telescope, NIKA2, and their successors in the field of dust polarization studies.

The primary deliverable of COSMOCal is the direct in-flight calibration of the polarized beams and absolute polarization angles of the participating observatories. These accurately calibrated instruments will then establish polarization standards based on the CMB and astrophysical sources such as the Crab Nebula, enabling the indirect calibration of future ground-based and space-based experiments. In this way, COSMOCal will establish a common absolute polarization reference, facilitating robust cross-experiment consistency and maximizing the scientific return of current and future polarization observations.

\section{IRAM 30m test campaign overview}
\label{sec:iram_cosmocal}
\begin{figure}[h!]
    \centering
    \includegraphics[width=0.7\textwidth]{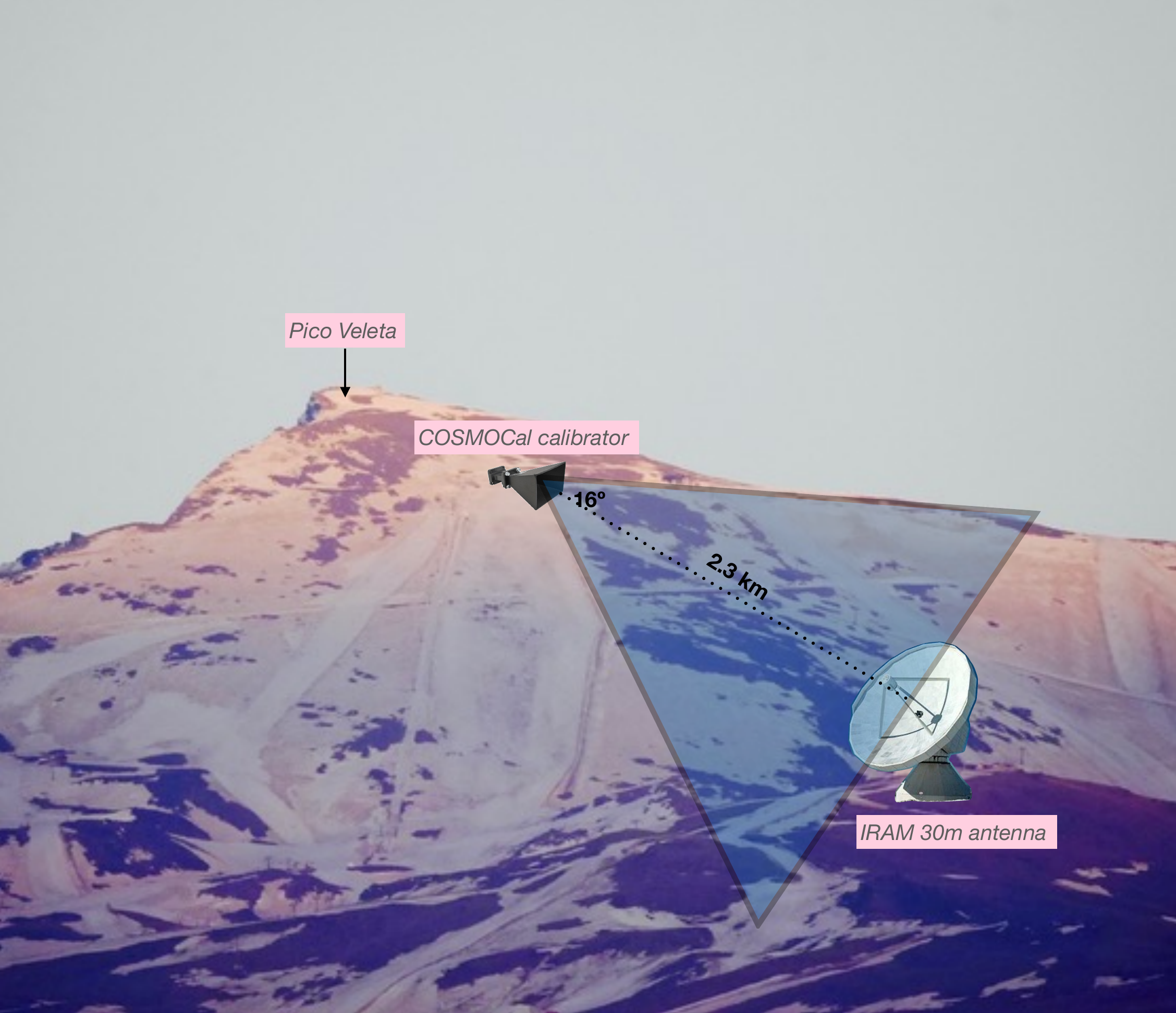}
    \caption{Schematic view of the IRAM 30~m telescope and COSMOCal source configuration during the Pico Veleta observing campaign.}
    \label{fig:scheme_picoveleta}
\end{figure}

To establish a proof of concept and validate the observational strategy that underlies the COSMOCal project, a prototype instrument has been developed to operate within the atmospheric window at 260~GHz. The COSMOCal prototype was equipped with a monochromatic source operating at 265~GHz, matching the polarization-sensitive channel of the NIKA2 camera (Adam et al. 2018\cite{adam2018}) installed on the IRAM 30~m telescope. The objective of this campaign was to demonstrate the performance of the calibrator coupled to a millimeter-wave telescope and to perform a first characterization of potential systematic effects arising from the telescope-calibrator optical chain. Although the measurements were conducted in the near field, they provide an important validation of the concept under realistic observing conditions. A future space mission currently under study with the CNES French space agency will enable far-field characterization of beam properties, polarization angle accuracy at 0.1º precision, ensuring cross calibration of ground observatories and future space missions such as LiteBIRD (Micheli et al 2026). NIKA2 was selected for its well-characterized polarization capabilities and high sensitivity (Perotto et al. 2020\cite{perotto2020}). The observations were carried out at the IRAM 30~m telescope on Pico Veleta, with the COSMOCal source installed at the University of Granada (UGR) site, approximately 100~m below the summit (Fig.~\ref{fig:scheme_picoveleta}). This location, previously used for telescope holography measurements, provided a direct line of sight to the antenna while simplifying deployment and operation.

Prior to deployment, simulations of the expected detector response were developed to support the analysis and optimize the observing strategy. The measurement campaign consisted of three stages: (i) detection of the COSMOCal signal, (ii) optimization of the telescope-source alignment through signal maximization, and (iii) polarization measurements obtained by rotating the source polarizer while performing fixed-track observations.

\section{Calibrator description}
\begin{figure}[h!]
    \centering
    \includegraphics[width=0.55\linewidth]{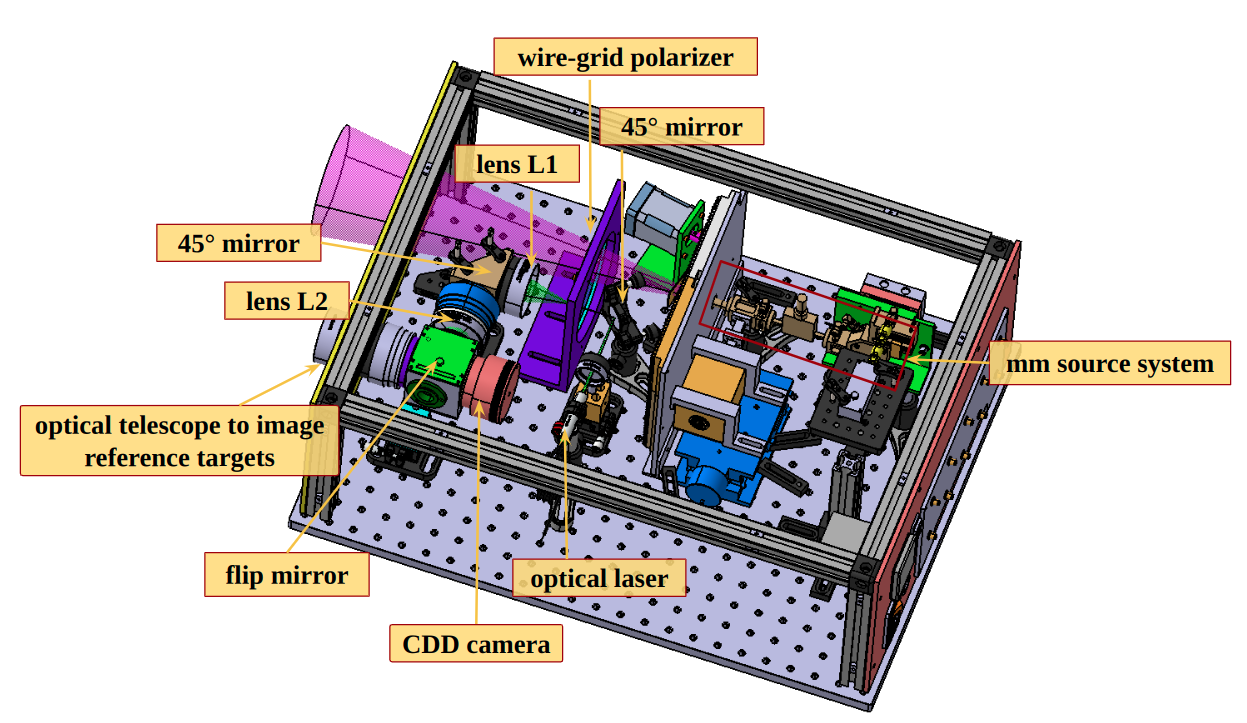}
    \includegraphics[width=0.44\linewidth]{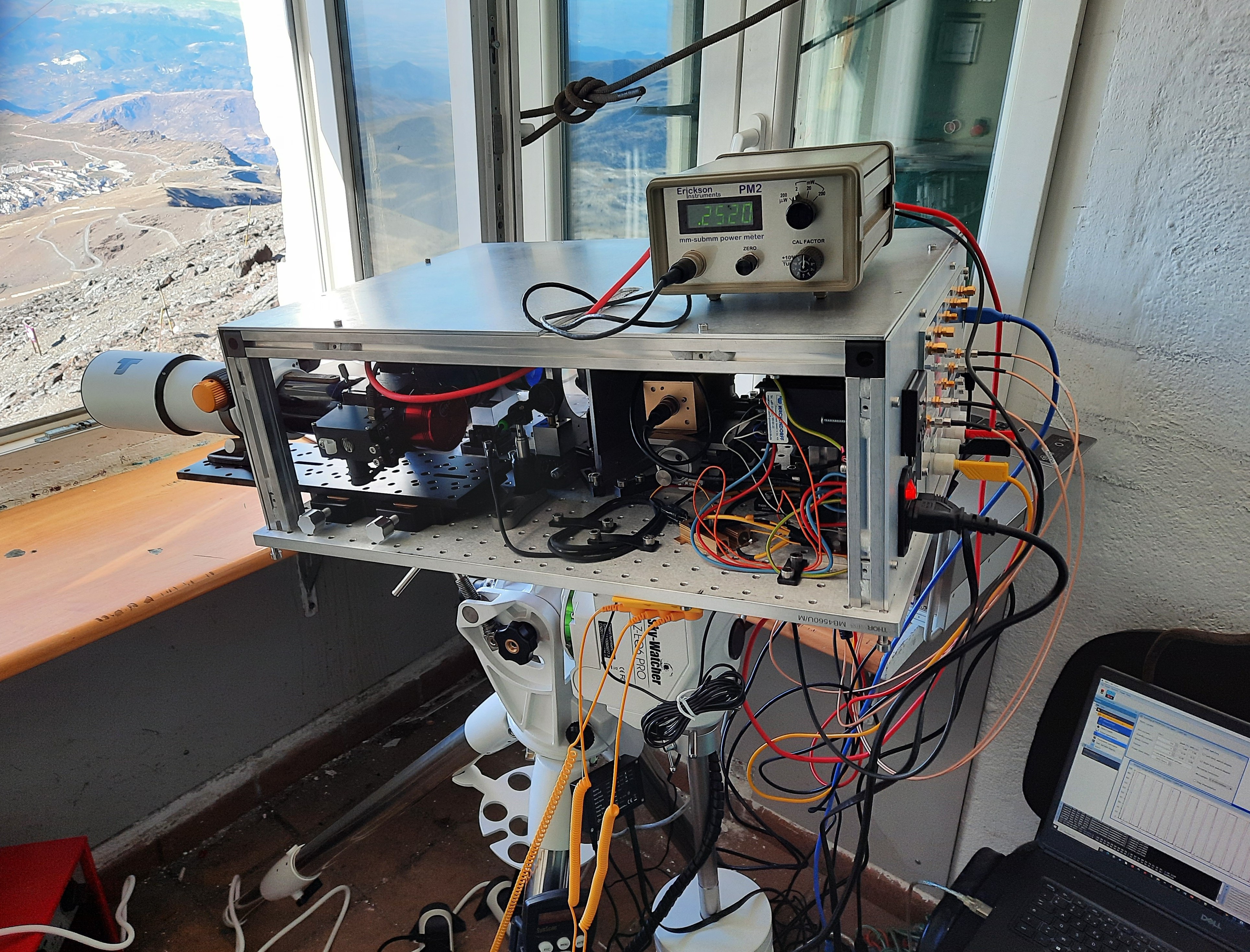}
    \caption{Left: COSMOCal design. Right: COSMOCal calibrator during the observing campaign at the IRAM 30m telescope.}
    \label{fig:box_pictures}
\end{figure}

The COSMOCal calibrator (Fig.~\ref{fig:box_pictures}) consists of a millimeter-wave emission chain and an optical metrology system that provides an independent reconstruction of the emitted polarization angle.

The millimeter source produces a $\sim 265$~GHz Gaussian beam with $\theta_{\mathrm{FWHM}} \simeq 20^\circ$. The output power is adjustable between $\sim 0.02$ and $\sim 2$~mW and the assembly is temperature-stabilized at $20^\circ$C using a thermoelectric system. A wire-grid polarizer placed at the horn output defines the polarization state and provides a coarse mechanical reference for the emitted polarization angle. An auxiliary optical system images both the polarization-dependent diffraction pattern of a reference laser and geo-referenced ground targets onto a CCD camera. The diffraction pattern orientation provides a direct proxy for the polarization direction in the camera frame, while the imaging of external targets enables an independent reconstruction of the calibrator attitude via photogrammetry. The combination of both methods allows a cross-validated determination of the emitted polarization angle with a target accuracy below $\sim 0.1^\circ$.

\section{Photogrammetry}
\label{sec:photogrammetry}
\begin{figure}[h!]
    \centering
    \includegraphics[width=0.8\linewidth]{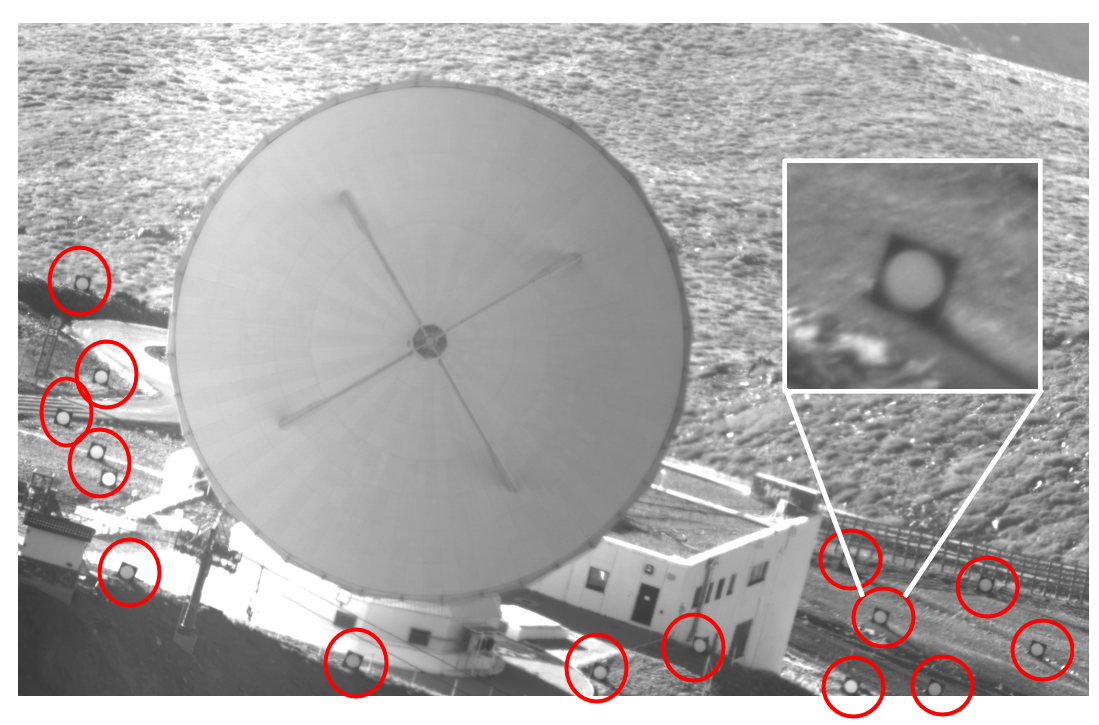}
    \caption{Ground targets observed by the COSMOCal CCD camera. Identified targets are highlighted in red.}
    \label{fig:target_highlight}
\end{figure}

Photogrammetry (Dünner et al. 2024\cite{dunner2024}) is used to reconstruct the attitude of the COSMOCal instrument with respect to the telescope line of sight. The method relies on imaging ground-based reference targets with a CCD camera mounted on the calibrator and solving a 3D--2D correspondence problem between known target positions and their projections in the camera plane (Fig.~\ref{fig:target_highlight}).

\subsection{Method overview}
Target positions are measured using RTK GPS and expressed in a local ENU frame centered on the NIKA2 cabin. Camera detections are obtained via a template-based finder, yielding target centroids with an accuracy of $\sim 2$ pixels, corresponding to $\sim 0.1$--$0.2$~m in the focal plane and compatible with GPS uncertainties at the site. The camera-to-world transformation is obtained using a PnP formulation as implemented in \textit{OpenCV}, solving for the rotation matrix and translation vector that minimize reprojection residuals. Camera intrinsic parameters and distortion coefficients are not fixed from laboratory calibration but re-estimated using a jackknife procedure to account for focus changes during on-site operation.

\subsection{Attitude reconstruction}
The calibrator attitude is derived by transforming the camera reference frame to the telescope line-of-sight frame through a composition of rotations,
\begin{equation}
R_{\mathrm{LOS}}^{c} = R_{w}^{c} \, R_{\mathrm{LOS}}^{w},
\end{equation}
where $R_{w}^{c}$ is obtained from the PnP solution and $R_{\mathrm{LOS}}^{w}$ is constructed geometrically from the telescope–source baseline. The resulting orientation is expressed in Euler angles (yaw, pitch, roll), with uncertainties estimated via jackknife resampling and propagation of target-position dispersion.

\subsection{Results}
The reconstructed yaw and pitch angles are consistent with a well-aligned configuration:
\begin{equation*}
\theta_{\mathrm{yaw}} = (-0.1672 \pm 0.0012)^\circ, \qquad
\theta_{\mathrm{pitch}} = (0.0271 \pm 0.0020)^\circ.
\end{equation*}

The roll angle exhibits a multimodal distribution when all target combinations are included, driven by sensitivity to target selection in the PnP solution. A stable solution is obtained by restricting to configurations with a sufficient number of detected targets and excluding degenerate geometries. Under these conditions, the calibrated roll angle is:
\begin{equation*}
\theta_{\mathrm{roll}} = (-15.1865 \pm 0.0308)^\circ.
\end{equation*}

This angle represents a fixed mechanical offset of the CCD reference frame with respect to the telescope line of sight and is used to correct the polarization reference frame in subsequent analysis. The dominant contribution to the uncertainty arises from GPS positioning accuracy in the installation environment.

\section{Diffraction pattern}
\label{sec:diffraction_pattern}
\begin{figure}[h!]
    \centering
    \includegraphics[width=0.75\linewidth]{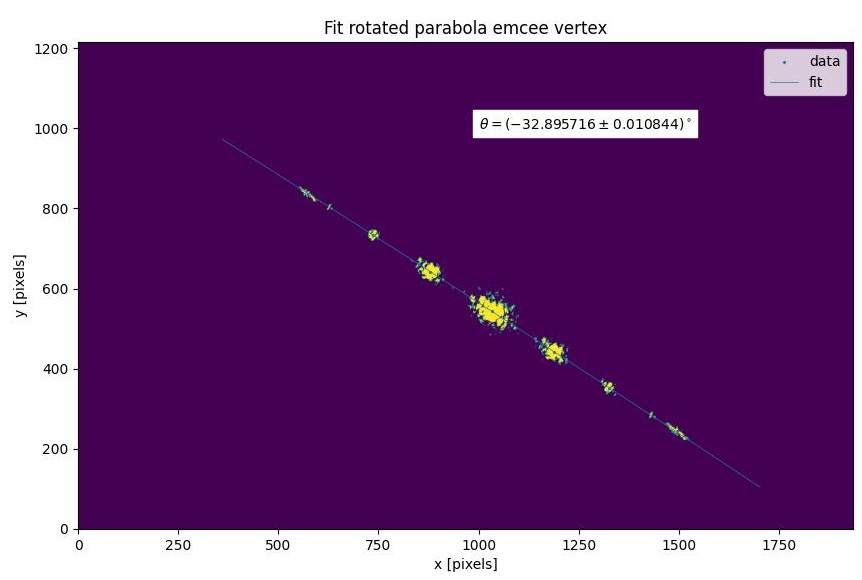}
    \caption{Diffraction pattern fit used to extract the polarization orientation in the camera plane. The rotated parabolic model is fitted to the centroid distribution of the diffraction maxima.}
    \label{fig:diffraction_pattern_images}
\end{figure}

The diffraction pattern provides a direct estimate of the polarization orientation in the camera reference frame. Combined with the photogrammetry-derived attitude, this enables projection of the measured polarization angle into the telescope line-of-sight frame, yielding an independent reconstruction of the emitted polarization as seen by NIKA2. The method is based on the analysis of the diffraction maxima produced by a laser beam passing through the polarizer. Their centroids $(x_{\mathrm{cen}}, y_{\mathrm{cen}})$ are modeled using a rotated parabolic structure. In the rotated reference frame, the transformation is written as
\begin{equation}
\begin{pmatrix}
x_{\mathrm{rot}} \\
y_{\mathrm{rot}}
\end{pmatrix}
=
R_{\theta}
\begin{pmatrix}
x_{\mathrm{cen}} - x_V \\
y_{\mathrm{cen}} - y_V
\end{pmatrix},
\end{equation}
with
\begin{equation*}
R_{\theta} =
\begin{pmatrix}
\cos\theta & \sin\theta \\
-\sin\theta & \cos\theta
\end{pmatrix},
\end{equation*}
where $(x_V, y_V)$ defines the vertex of the parabola. In this frame, the diffraction structure is described by $y_{\mathrm{model}} = a x_{\mathrm{rot}}^2$. The parameters $(x_V, y_V, \theta, a)$ are inferred via an MCMC procedure that minimizes residuals between model and observed centroids. The angle $\theta$ corresponds to the polarization orientation in the camera plane and is subsequently corrected using the photogrammetry solution to obtain the orientation in the telescope reference frame. A representative diffraction pattern image and the corresponding fit are shown in Fig.~\ref{fig:diffraction_pattern_images}.

\section{Optical model}
\label{sec:opt_mod}
To characterize the propagation of the COSMOCal signal in a near-field configuration and assess the impact of source misalignment on the illumination of the NIKA2 focal plane, we developed a simplified optical model based on ray tracing and Gaussian beam propagation. The model is also used to cross-check expected power levels at detector level and ensure operation within the linear regime of the KIDs.

\subsection{Near-field propagation and misalignment effects}
\begin{figure}[h!]
    \centering
    \includegraphics[width=0.75\textwidth]{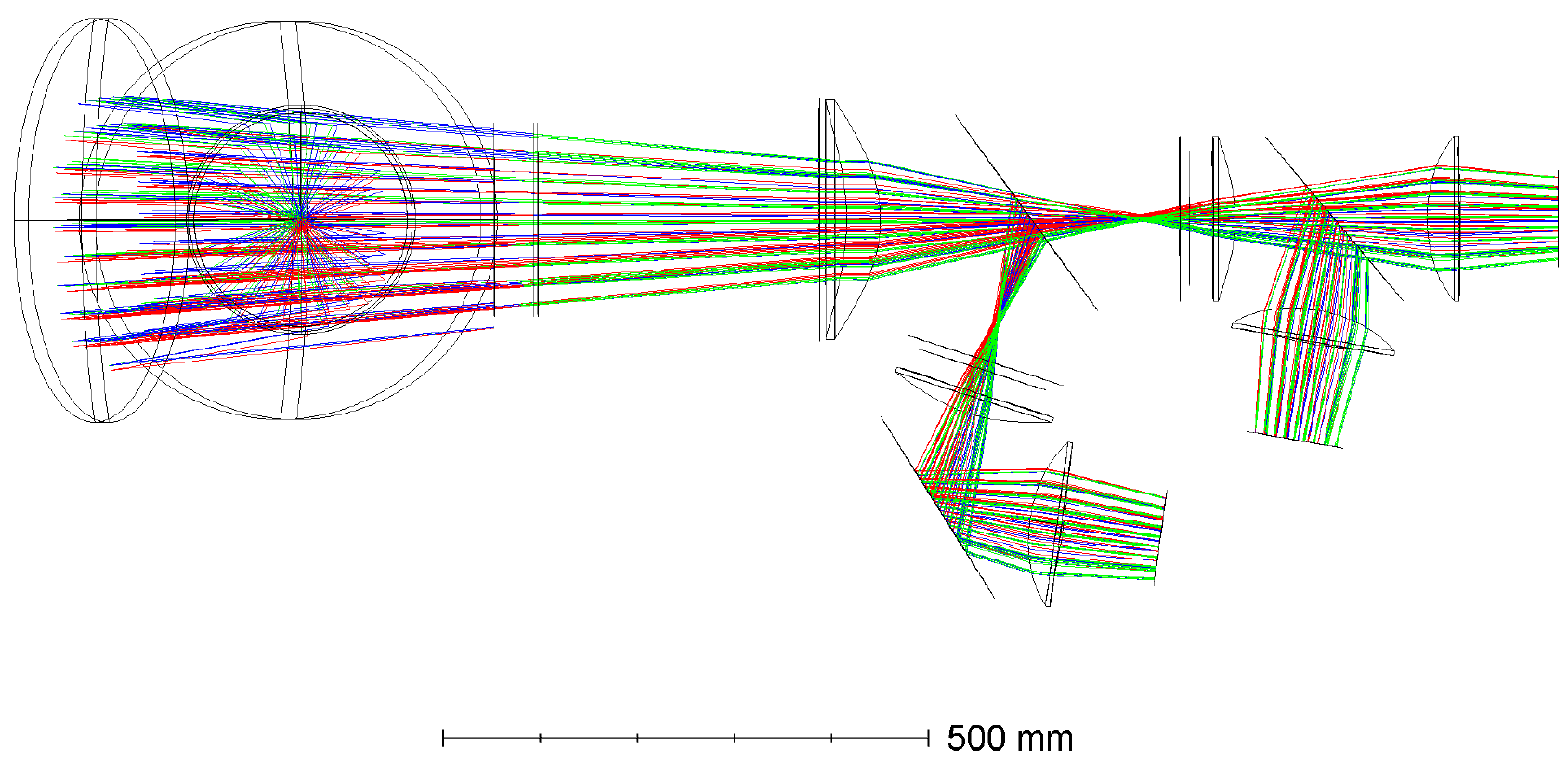}
    \caption{Ray propagation from the telescope entrance to the focal plane for different source misalignments. Blue corresponds to nominal alignment, while red and green illustrate increasing lateral offsets. Only unvignetted rays reaching the detector plane are shown.}
    \label{fig:rays_compare}
\end{figure}

The source is located at a distance of 2.31~km from the IRAM 30~m telescope and emits a Gaussian beam with $\sim16^\circ$ FWHM. In this regime, the system operates in a near-field configuration, and the illumination of the primary mirror is not well approximated by plane-wave propagation. We explore the effect of lateral pointing offsets between source and telescope optical axis, which modify the overlap between the Gaussian beam and the telescope pupil. Small offsets can partially reduce central obscurations from the secondary mirror and support structures, while larger offsets progressively vignette the pupil until the source leaves the effective field of view. A representative ray-tracing comparison is shown in Fig.~\ref{fig:rays_compare}, illustrating the evolution of the optical coupling for increasing misalignment.

Despite geometric simplifications, the model captures the key behaviour: an initial regime where illumination improves as central obstructions are partially avoided, followed by a monotonic decrease in coupling as vignetting dominates.

\subsection{Power scaling estimate}
At 2.3~km distance, the illuminated radius of a $16^\circ$ beam is approximately
\begin{equation}
r \simeq 2300 \times \tan(8^\circ) \approx 323~\mathrm{m}.
\end{equation}

This leads to a strong geometric dilution between the source illumination scale and the telescope collecting area, with an effective factor
\begin{equation}
D \sim 4.5 \times 10^{-5}.
\end{equation}

Assuming an emitted power of 0.2~mW and an overall optical efficiency of 35\% (Adam et al. 2018\cite{adam2018}), the expected power per detector (for $\sim10^3$ pixels) is
\begin{equation}
P_{\mathrm{pix}} \sim 3~\mathrm{pW}.
\end{equation}

This value is consistent with the expected dynamic range of the NIKA2 KIDs and ensures operation below saturation.

\subsection{On-site calibration and measured response}
The emitted power was monitored via a calibrated directional coupler placed before the horn, which splits the 1\% of the power in a power meter, allowing continuous tracking of the transmitted power during operation. A power level of order 1~mW was found to saturate the detectors, and the operating point was therefore reduced to $\sim0.2$~mW. Using Neptune as a calibration source, we estimate a single-KID responsivity of
\begin{equation}
\mathcal{R} \simeq 0.4~\mathrm{pW/kHz}.
\end{equation}

For a measured frequency shift of $\sim 40$~kHz induced by the COSMOCal source, the inferred received power per detector is
\begin{equation}
P_{\mathrm{KID}} \sim 15~\mathrm{pW}.
\end{equation}

This is higher than the simplified optical estimate but remains compatible within uncertainties associated with near-field coupling, beam truncation effects, and model approximations. Importantly, it confirms that the detectors remain in a linear and stable operating regime.

\section{Signal model and preliminary detection}
\label{subsec:sig_model_iram}
The expected signal modulation is described using a Mueller matrix formalism including the COSMOCal polarizer, the rotating half-wave plate (HWP) of NIKA2, and the cold analyzer polarizer. Starting from an unpolarized input state, the detected intensity can be written as

\begin{equation}
S_{\mathrm{out}} = M_{\mathrm{pol}}(\beta)\, M_{\mathrm{HWP}}(\theta)\, M_{\mathrm{pol}}(\psi)\, S_{\mathrm{in}},
\end{equation}

where $\psi$ is the COSMOCal polarizer angle, $\theta=\omega t$ is the HWP rotation angle, and $\beta$ is the analyzer orientation.

This leads to a modulated signal of the form

\begin{equation}
S(t) = 1 + \cos\!\left[4\theta - (2\beta + 2\psi)\right],
\end{equation}

showing the characteristic $4f_{\mathrm{HWP}}$ modulation expected from a rotating half-wave plate system. An additional square-wave modulation is introduced by an external chopper at frequency $f_{\mathrm{chop}}$, producing a combined amplitude-modulated signal used for synchronous detection.

\subsection{Preliminary detection and validation}
\begin{figure}[h!]
    \centering
    \begin{minipage}{0.49\textwidth}
        \centering
        \includegraphics[width=\linewidth]{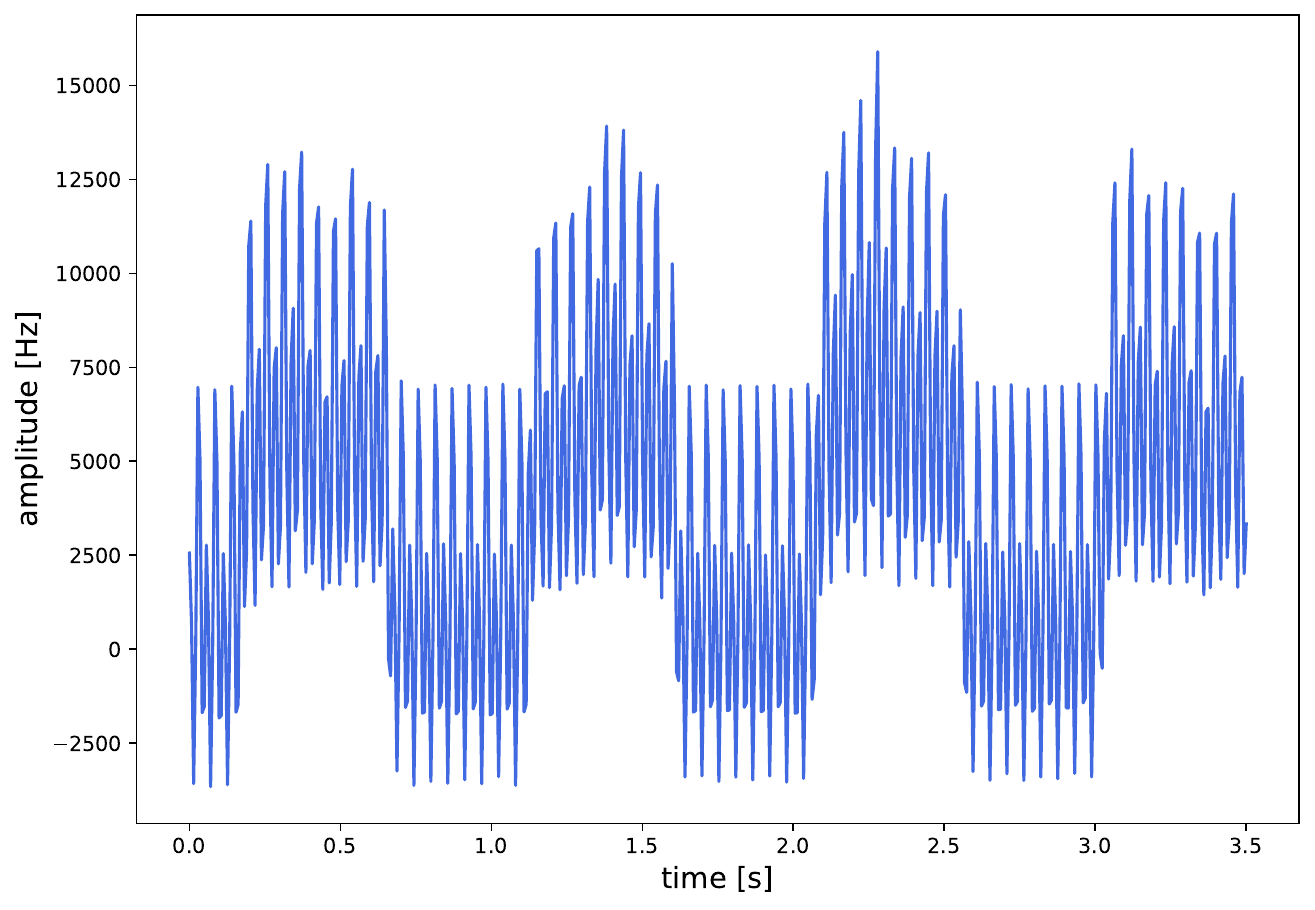}
    \end{minipage}\hfill
    \begin{minipage}{0.49\textwidth}
        \centering
        \includegraphics[width=\linewidth]{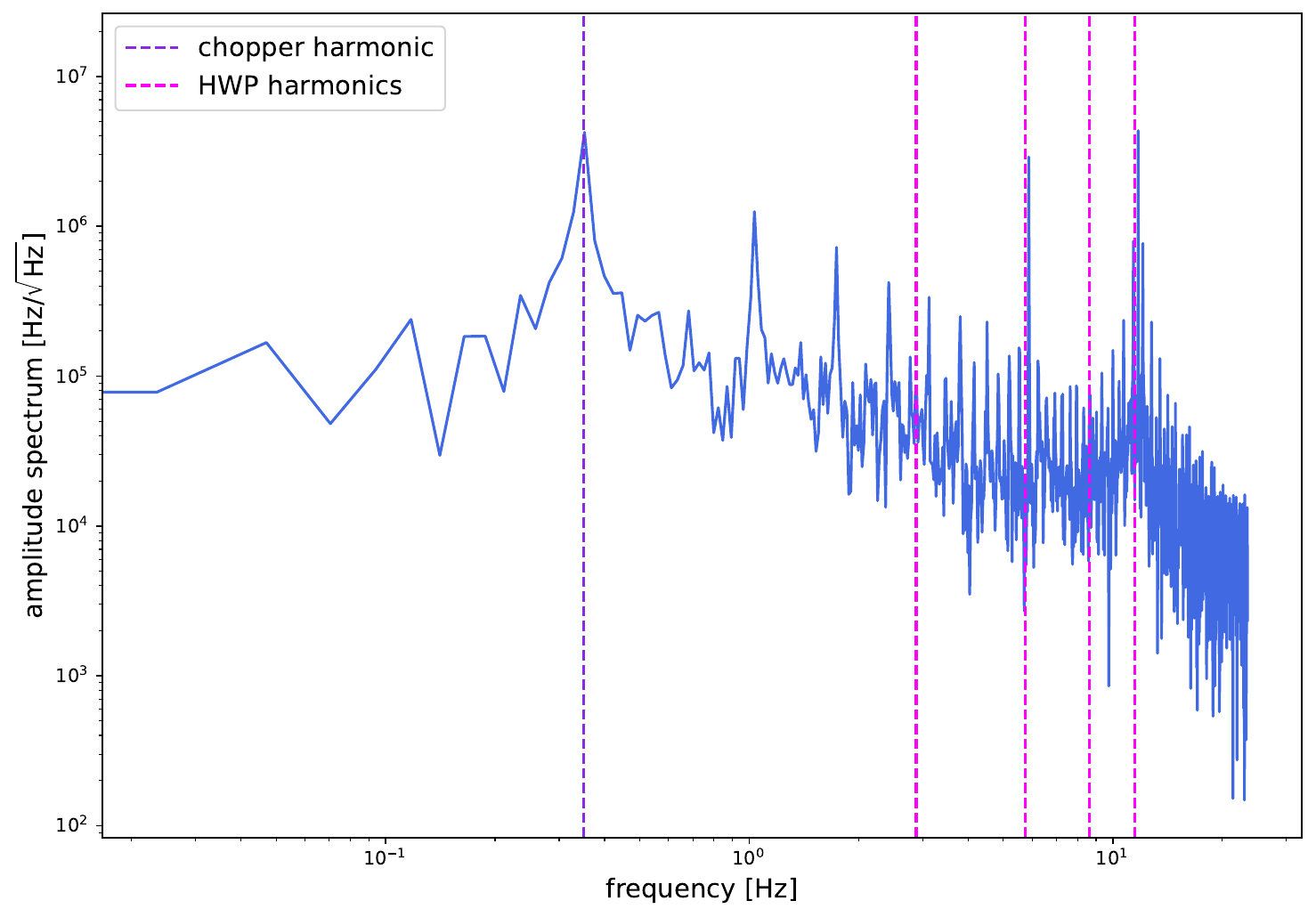}
    \end{minipage}
    \caption{Timeline of one sample pixel of A1 (\textit{left}) and relative power spectrum (\textit{right}). The dashed lines in violet the chopper's main harmonic and in pink the 4 first harmonics of the HWP.}
    \label{fig:samp_timeline_pws}
\end{figure}

The first on-sky tests aimed at verifying the presence of both modulation signatures in the detector timestreams. Despite varying atmospheric and background conditions, the COSMOCal source produces a clear response in NIKA2 detectors. A representative timestream and corresponding power spectrum are shown in Fig.~\ref{fig:samp_timeline_pws}. The signal exhibits the expected ON/OFF structure from the chopper and a stable harmonic content associated with the rotating HWP.

The observed spectral peaks occur at the expected frequencies, confirming correct operation of the modulation chain and stable detection of the COSMOCal signal. Deviations from the ideal simulated waveform are observed in the timestream amplitude, consistent with partial depolarization effects and non-ideal transmission of the HWP, as well as additional background contributions in the near-field observational setup.

Overall, these results demonstrate successful recovery of both modulation signatures and establish the basis for subsequent polarized signal extraction.

\section{Pipeline for data reduction: key concepts}
\label{sec:pipeline}

We briefly summarize the data analysis pipeline used to extract intensity and polarization information from raw NIKA2 KID timelines. The goal is to reconstruct the Stokes parameters $I$, $Q$, and $U$ while separating astrophysical signal from instrumental and atmospheric contributions.

The raw timelines include the COSMOCal signal, atmospheric fluctuations, and instrumental polarization induced primarily by the rotating half-wave plate (HWP). The detector signal can be modeled as

\begin{equation}
S(t) = I + Q \cos(4\phi_{\mathrm{HWP}}) + U \sin(4\phi_{\mathrm{HWP}}),
\end{equation}

where $\phi_{\mathrm{HWP}}$ is the instantaneous HWP angle derived from the synchronization signal.

A key challenge is the presence of a parasitic HWP synchronous signal (HWPSS), which adds harmonic contamination at multiples of the rotation frequency in the Fourier space. This component, together with slow drifts and atmospheric fluctuations, is removed through an iterative procedure that exploits the chopper modulation. The chopper signal provides a reference for identifying on/off source plateaus, allowing separation of source-dominated and background-dominated intervals. Instrumental templates for the HWPSS are then fitted on selected low-signal intervals and subtracted from the full timeline, followed by baseline removal to correct residual drifts.

After cleaning, the data are projected onto the $\cos(4\phi_{\mathrm{HWP}})$ and $\sin(4\phi_{\mathrm{HWP}})$ modes to estimate $Q$ and $U$ for each detector and each scan. The results are then averaged over stable plateaus and across detectors to improve signal-to-noise.

From the reconstructed Stokes parameters, we compute the polarized intensity and polarization angle as

\begin{equation}
I_{\mathrm{pol}} = \sqrt{Q^2 + U^2}, \qquad
\psi = \frac{1}{2}\arctan(U/Q),
\label{eq:polang}
\end{equation}

with polarization fraction $p = I_{\mathrm{pol}}/I_{\mathrm{tot}}$.

This pipeline is applied independently to the two NIKA2 arrays (A1 and A3), enabling cross-checks and providing robust estimates of both total intensity and polarized signal. The combination of chopper demodulation and HWP harmonic analysis ensures efficient separation of source signal from instrumental and atmospheric contamination.

\section{Radio alignment}
\label{sec:radio_alignment}
\begin{figure}[h!]
    \centering
    \begin{minipage}{0.49\textwidth}
        \centering
        \includegraphics[width=\linewidth]{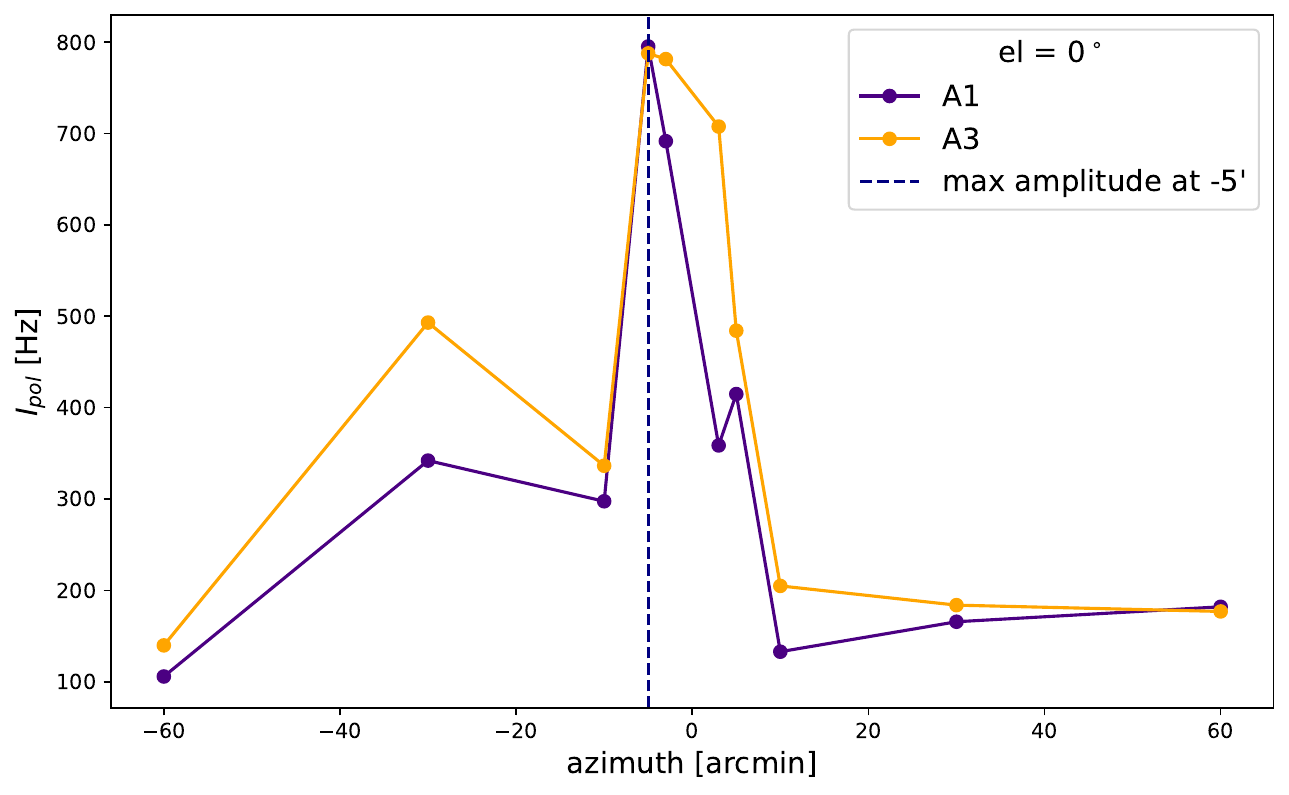}
    \end{minipage}\hfill
    \begin{minipage}{0.49\textwidth}
        \centering
        \includegraphics[width=\linewidth]{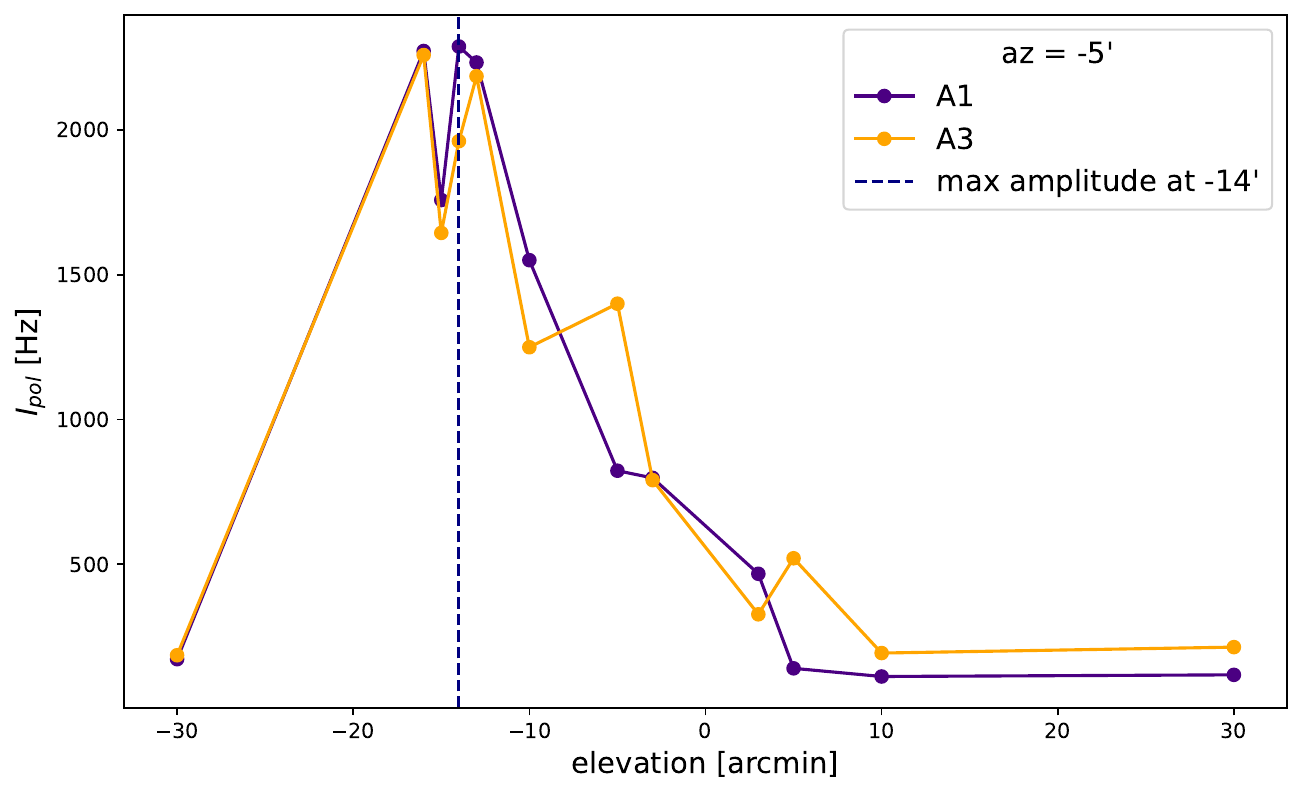}
    \end{minipage}
    \caption{Polarized intensity as a function of azimuth (left) and elevation (right) pointing offsets. The maxima define the optimal alignment corrections.}
    \label{fig:Ipol_vs_azel}
\end{figure}

Accurate pointing of the COSMOCal source relative to the IRAM 30~m antenna was required to maximize coupling efficiency, as suggested by the optical simulations discussed in Sec.~\ref{sec:opt_mod}. Starting from the nominal pointing coordinates (AzEl = [116.19º, 10.49º]), a dedicated scan campaign was performed to refine the alignment. The procedure consisted in sequential small-offset scans in azimuth and elevation, using the reconstructed polarized intensity as a figure of merit. First, an azimuthal scan was performed at fixed elevation, yielding a best-fit correction of $-5'$ corresponding to the maximum signal response. Subsequently, an elevation scan was performed at fixed corrected azimuth, resulting in an additional optimal offset of $-14'$. The resulting optimization curves are shown in Fig.~\ref{fig:Ipol_vs_azel}, where the polarized intensity is displayed as a function of the applied pointing corrections in both axes.

The final pointing correction is therefore $(-5', -14')$ in azimuth and elevation, respectively, leading to updated coordinates [116.10º, 10.25º] for subsequent observations.

\section{Polarization measurements}
\label{sec:pol_measurements}
After alignment, polarization measurements were performed by rotating the COSMOCal polarizer to different orientations and acquiring fixed-track scans for each configuration. The polarization angles used in the analysis are those independently reconstructed combining photogrammetry results with diffraction pattern analysis (Sec.~\ref{sec:diffraction_pattern}). All angles are expressed in the AzEl convention.

\subsection{Verification of Malus law}
\label{subsec:malus}
\begin{figure}[h!]
    \centering
    \begin{minipage}{0.5\textwidth}
        \centering
        \includegraphics[width=\linewidth]{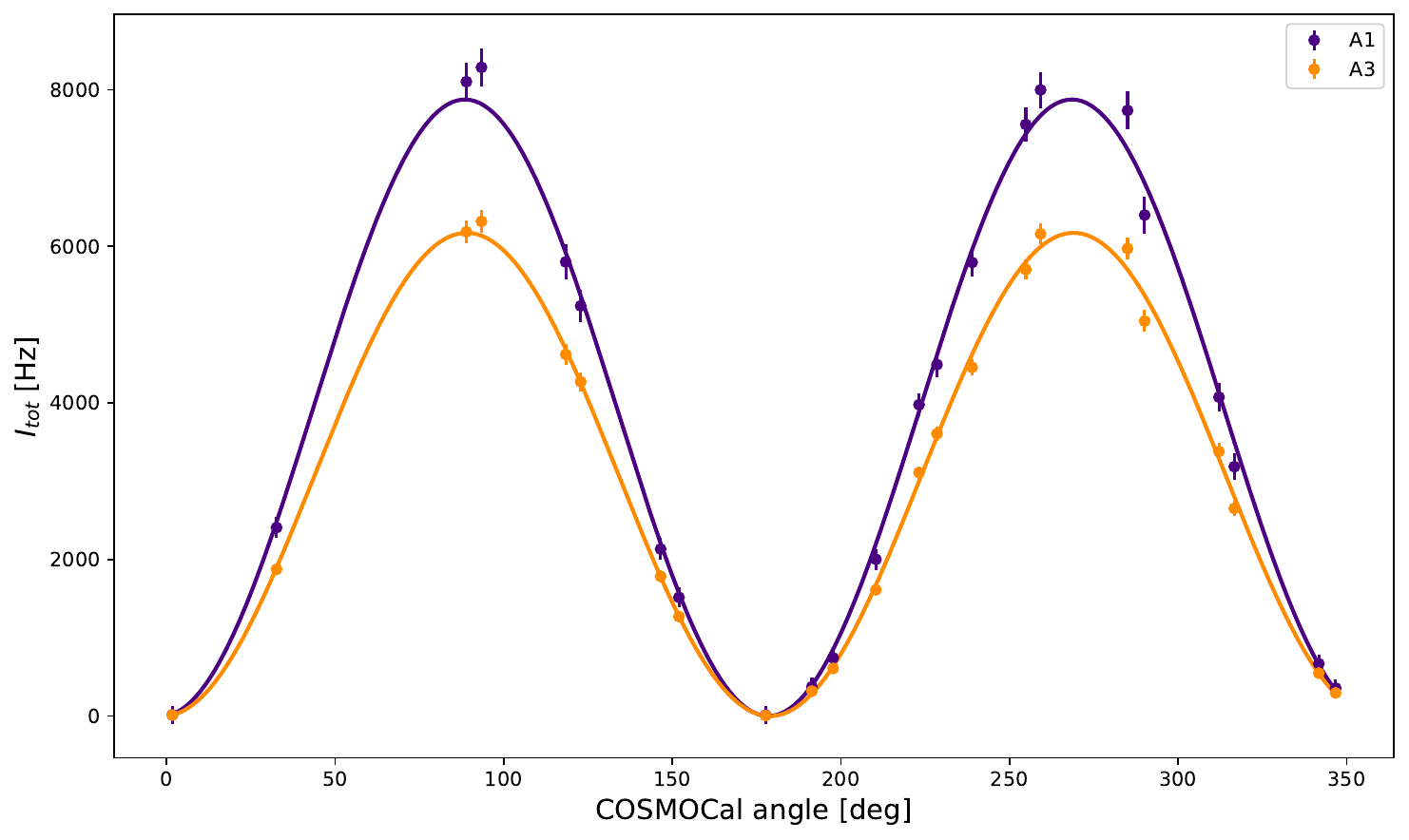}
    \end{minipage}\hfill
    \begin{minipage}{0.5\textwidth}
        \centering
        \includegraphics[width=\linewidth]{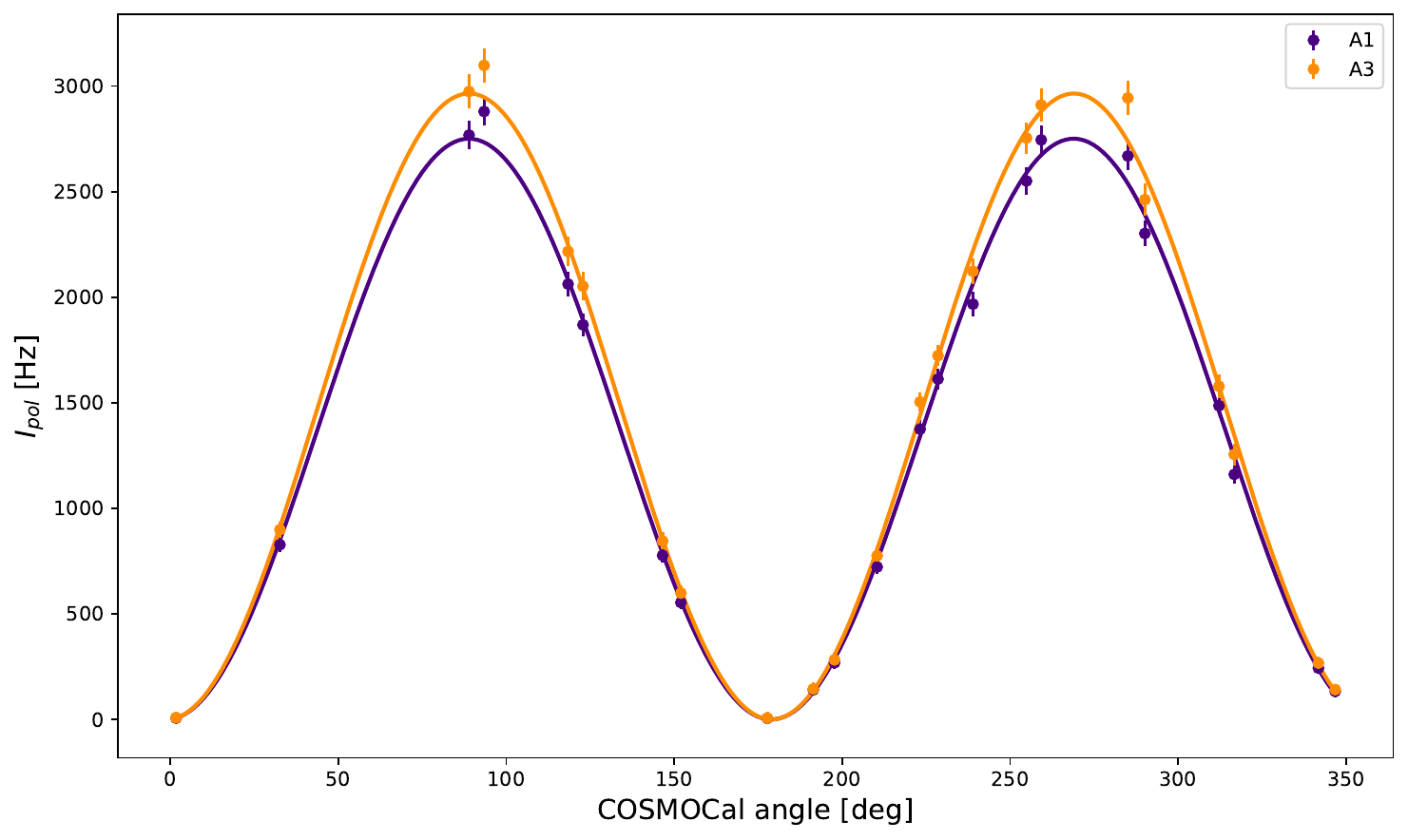}
    \end{minipage} 
    \caption{\textit{Left}: average total intensity (Stokes I parameter) as a function of the rotational angle of COSMOCal's polarizer. \textit{Right}: average polarized intensity ($\sqrt{Q^2+U^2}$) again as a function of COSMOCal's polarizer rotation angle.}
    \label{fig:malus}
\end{figure}

As a first consistency check, we verify Malus’ law on both total and polarized intensity. For an ideal polarizer, the transmitted intensity follows
\begin{equation}
I = I_0 \cos^2(\varphi),
\end{equation}
where $\varphi$ is the relative angle between the incoming polarization and the polarizer axis. The measured intensity modulation as a function of the COSMOCal polarizer angle is well described by this dependence, with a best-fit phase close to $90^\circ$, as shown in Fig.~\ref{fig:malus}, consistent with the expected geometrical configuration. A slight difference between arrays A1 and A3 is observed in the absolute intensity scale, while the angular modulation remains consistent.

The polarization degree is found to be below unity and shows a mild dependence on the array, indicating partial depolarization effects likely related to near-field propagation and instrumental transfer effects.

\subsection{Polarization angle reconstruction}
The absolute polarization angle of the emitted signal is obtained by combining diffraction pattern reconstruction (camera-plane orientation) with photogrammetry (roll angle):
\begin{equation}
\psi = \frac{\pi}{2} - \theta_{\mathrm{dp}} + \theta_{\mathrm{roll}}.
\end{equation}

The resulting independent estimate of the polarization angle is consistent with the expected mechanical orientation of the polarizer within approximately 2.6$^\circ$. This discrepancy is likely due to uncertainties in the manual alignment of the polarizer, implying that the actual wire orientation may differ from that reconstructed from the diffraction pattern and photogrammetry.
The final uncertainty is dominated by the roll angle reconstruction, leading to a typical precision of order

\begin{equation}
\sigma_\psi \sim 0.03^\circ,
\end{equation}

for the internal COSMOCal determination, while the dominant limitation in the comparison with NIKA2 remains the telescope-side systematic budget.

\begin{figure}[h!]
    \centering
    \includegraphics[width=0.75\textwidth]{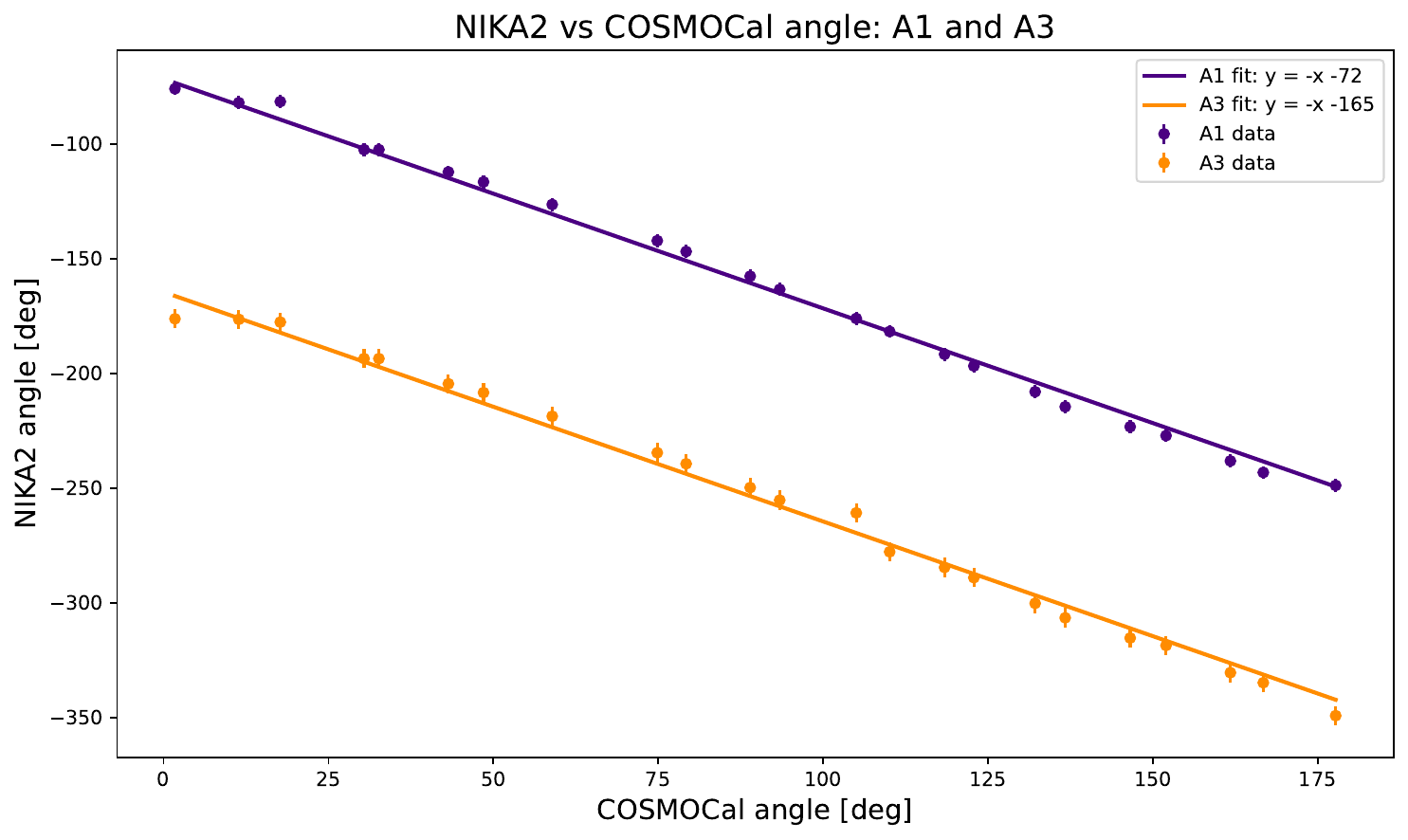}
    \caption{Values of the polarization angle $\psi$ detected by NIKA2 as a function of the angles found by the internal independent COSMOCal's system.}
    \label{fig:pol_angle_cal}
\end{figure}

The main scientific goal is the reconstruction of the polarization angle as seen by NIKA2 and its comparison with the independent COSMOCal reference. For each polarizer orientation, Stokes parameters are averaged over each array, and the polarization angle is derived as in Eq.~\ref{eq:polang}. A clear linear relationship is observed between the COSMOCal reference angle and the reconstructed NIKA2 angle for both arrays (Fig.~\ref{fig:pol_angle_cal}), confirming the expected geometrical response of the instrument. The slope is consistent with $-1$ as expected from the optical configuration, and the intercept encodes the residual global offset of the system.

Residual discrepancies between arrays are observed at the level of a few degrees, which are discussed in terms of instrumental systematics and residual calibration uncertainties.

\subsection{Uncertainty budget}
The dominant contributions to the polarization angle uncertainty arise from (i) focal plane dispersion, (ii) residual systematics across the arrays, and (iii) mechanical tolerance in the polarizer angle setting. Combining these effects, the overall uncertainty on the polarization angle is estimated to be of order
\begin{equation}
\sigma_\psi \sim 3^\circ,
\end{equation}
which is significantly larger than the sub-degree target but consistent with a near-field configuration and non-ideal optical coupling.

\section{Conclusions}
The COSMOCal IRAM campaign demonstrates a consistent reconstruction of polarization observables through two independent and complementary approaches: the polarization angle inferred from the calibrator geometry, reconstructed using diffraction pattern analysis and photogrammetry, and the angle measured directly by NIKA2 detector polarimetry. The agreement between these methods confirms the internal coherence of the system and validates the overall calibration strategy in a realistic observational environment.

The independent calibration capability of COSMOCal is successful, but its performance is still limited by identifiable instrumental constraints. On the photogrammetry side, the restricted field of view of the optical system limits the number and geometry of ground reference targets, weakening the conditioning of the camera-to-world transformation and increasing the dispersion of the reconstructed attitude. A wider field-of-view system and a denser, more stable target distribution would significantly improve the accuracy of the reconstruction. In addition, both photogrammetry and diffraction pattern reconstruction would benefit from higher-statistics acquisition strategies, such as continuous imaging, which would reduce centroiding noise and improve robustness.

Despite these limitations, the calibrator achieves a reliable independent polarization reconstruction consistent with the expected performance for ground validation of the methodology. The combined diffraction and photogrammetry approach provides a viable path toward an absolute polarization reference at the required level of accuracy, even though the present implementation remains statistically and systematically limited.

On the NIKA2 side, the expected $\sim 90^\circ$ relative orientation between arrays A1 and A3 is recovered, and a clear linear correlation is observed between the COSMOCal reference angles and the reconstructed detector polarization angles. This confirms that the instrument responds correctly to controlled polarization modulation. However, a spatial gradient of polarization angle across the focal plane is observed, which complicates the interpretation of statistical uncertainties and introduces an additional systematic component. This effect is consistent with previously known instrumental behavior associated with the dichroic element.

A second major limitation arises from the near-field configuration of the experiment. The extended-source geometry, combined with residual optical reflections and non-ideal propagation effects, leads to a reduction in the measured polarization degree relative to the ideal expectation. This cannot be explained solely by the finite transmission of the half-wave plate and indicates additional depolarization mechanisms intrinsic to the setup. Together, these effects currently dominate the uncertainty budget and prevent reaching the target precision of $0.1^\circ$ on the polarization angle.

The resulting total uncertainty, physically interpreted in terms of these two dominant systematics, is estimated to be of the order of $3^\circ$. Both limitations are, however, addressable. The dichroic-related systematics are expected to be strongly reduced following the NIKA2+ upgrade (March 2025), which replaced the dichroic filter. The near-field limitation requires a different experimental configuration, either with a significantly larger source distance or with an alternative setup more representative of far-field conditions.

Overall, this campaign provides a clear demonstration of the main systematic effects affecting ground-based polarimetric calibration, while simultaneously validating the core principles of the COSMOCal approach. The results establish a solid foundation for the next phase of development, namely an upgraded prototype closer to the final space instrument concept. This future system will incorporate improved optical focusing and a controlled pointing mechanism, enabling structured illumination of the focal plane and systematic characterization of instrumental polarization response.

In the longer term, the results support the feasibility of a space-based COSMOCal-like calibrator. In such a configuration, the need for photogrammetric reconstruction may be relaxed, provided that spacecraft pointing and instrument geometry are independently controlled at sufficient precision. The IRAM campaign therefore represents both a validation of the methodology and a guide for simplifying and optimizing the system design for future deployments.

\section*{Acknowledgments}
We acknowledge financial support from CENSUS, Observatoire de Paris-PSL and CNES space agency. A. Ritacco acknowledges financial support from the Italian Ministry of University and Research - Project Proposal CIR01\_00010. F. Nati acknowledges funding from the European Union (ERC, POLOCALC, 101096035).

\bibliographystyle{spiebib}  
\bibliography{report} 

\end{document}